\documentclass[twocolumn,superscriptaddress,showpacs,preprintnumbers,prb,amsmath,amssymb]{revtex4}

\usepackage{graphicx}
\usepackage{dcolumn}
\usepackage{bm}

\begin{document}

\title{Spontaneous ferromagnetic spin ordering at the surface of La$_2$CuO$_4$}

\author{R. V. Yusupov}
\altaffiliation[Permanent address: ]{Kazan State University,
Kremlevskaya 18, 420008 Kazan, Russia} \email[E-mail:
]{Roman.Yusupov@ijs.si}  \affiliation{Jozef Stefan Institute, Jamova
39, 1000 Ljubljana, Slovenia}

\author{V. V. Kabanov}
\affiliation{Jozef Stefan Institute, Jamova 39, 1000 Ljubljana,
Slovenia}

\author{D. Mihailovic}
\affiliation{Jozef Stefan Institute, Jamova 39, 1000 Ljubljana,
Slovenia} \affiliation{University of Ljubljana, Jadranska 19, 1000
Ljubljana, Slovenia}

\author{K. Conder}%
\affiliation{Laboratory for Neutron Scattering, ETH Z\"{u}rich and
PSI, CH-5232 Villigen PSI, Switzerland}

\author{K. A. M\"{u}ller}
\affiliation{Physik-Institut der Universit\"{a}t Z\"{u}rich,
Winterthurerstrasse 190, CH-8057, Switzerland}

\author{H. Keller}%
\affiliation{Physik-Institut der Universit\"{a}t Z\"{u}rich,
Winterthurerstrasse 190, CH-8057, Switzerland}

\date{\today}

\begin{abstract}
Magnetic properties of high purity stoichiometric La$_2$CuO$_4$ nanoparticles are systematically investigated as a function of particle size. Ferromagnetic single-domain spin clusters are shown to
spontaneously form at the surface of
fine grains as well as paramagnetic defects. Hysteresis loops and thermomagnetic irreversibility are
observed in a wide temperature range $5 - 350$ K with the remnant
moment and coercivity gradually decreasing with increasing temperature.
 Possible origins of the spontaneous surface ferromagnetic clusters and the
relation of our data to the appearance of unusual magnetic phenomena and phase separation of doped cuprates are
discussed.

\end{abstract}


\maketitle

\section{Introduction}
High-temperature superconductors (HTSC)
have been investigated for more than twenty years but still new important
details of their physical structure and properties are being discovered.
Moreover, in spite of the clear progress achieved in the
clarification of the phase diagram of cuprates,
the spin dynamics in relation to superconductivity in these
compounds remains unclear.

One of the things which is no doubt of crucial importance in undoped and lightly doped cuprates
is magnetism and magnetic (exchange)
interactions. This is clear from the simple observation that the
magnetic order in La$_2$CuO$_4$ is strongly influenced by the
non-stoichiometry or chemical doping needed for the
superconductivity to arise (see e.g.
Ref.\onlinecite{Kastner_RMP_98}). Thus, the parent La$_2$CuO$_4$ has
the N\`{e}el temperature $T_N = 325$ K. For La$_{2-x}$Sr$_x$CuO$_4$
$T_N$ decreases sharply with the increase of Sr doping. In the case
of La$_2$CuO$_{4+y}$ the situation is even more peculiar. The
compound within the so-called miscibility gap (1\% - 6\% of excess
oxygen) tends to phase separate into the superconducting (below
$\sim 40$ K) oxygen-rich and nearly-stoichiometric oxygen-poor
regions. The sample is found to be macroscopically inhomogeneous,
and the N\`{e}el temperature for the oxygen-poor phase is $T_N
\approx 260$ K being the characteristic one in the shown above
rather wide range of the excess oxygen concentrations. It looks
clear that in the case of Sr-doped La$_2$CuO$_4$ the local
inhomogeneity due to impurity disorder and structural twinning is
also the intrinsic property of the material.

The unusual magnetic properties of the fine grains and nanoparticles
of the antiferromagnetic (AF) in the bulk transition metal oxides
were predicted by N\`{e}el \cite{Neel} to arise from the
uncompensated magnetic moments of the same sublattice found at the
grain surface. Such magnetism was really observed and attracted the
attention of the researchers due to possible its practical use.
During the last decades the results on NiO
\cite{Richardson_NiO1,Cohen_JPSJ_62,Richardson_NiO2,
Makhlouf_NiO_JAP, Kodama_PRL_NiO}, MnO \cite{Sako_JPSJ_MnO},
Cr$_2$O$_3$ \cite{Cohen_JPSJ_62}, CoO \cite{Flipse_EurPhysJB99,
Zhang_JMMM03, Ghosh_ChemMat05}, Fe$_2$O$_3$ \cite{Cohen_JPSJ_62},
CuO \cite{Punnoose_PRB_CuO} and ferritin \cite{Tejada_JPCM,
SGider04071995, Makhlouf_PRB_97} have been published. Large magnetic
moments were found and phenomena peculiar, especially for
antiferromagnets were observed, such as superparamagnetism and
exchange bias.

In this paper we present the results of the detailed magnetic
studies of stoichiometric La$_2$CuO$_4$ fine grains. The compound
has a layered perovskite crystal structure and its magnetic
structure is more complicated than the structures of binary
transition metal oxides. It is almost ideal 2D Heisenberg
antiferromagnet with the strong superexchange in the CuO$_2$ planes.
Spin canting due to Dzyaloshinskii-Moriya interaction below the
structural phase transition at $\sim 530$ K produces a
non-zero out-of-plane magnetic moment for each CuO$_2$ plane. These
moments order antiferromagnetically at $T_N = 325$ K. The interaction of
the moments with an applied magnetic field can overcome the
interplane exchange interaction and lead to the so-called weak
ferromagnetism \cite{Cheong_PRB_89}.

We have found an unusual relatively strong nonlinear component in
its magnetization as a function of applied field characteristic of
anisotropic ferromagnetic single-domain particles. It is shown that
this magnetism arises from the surface of the material, but  is not
due to uncompensated surface moments proposed by N\`{e}el. Grain
boundaries are known as one of the limiting factors for practical
applications of HTSC materials and the surface phenomenon described
here may be important from this point of view. Some of the
hysteretic behavior of the magnetization reported for cuprates in a
number of works \cite{Kremer_ZfP_92, Kremer_ZfP_93, Sigmund_ZfP_94,
panagopoulos:144508, majoros:024528, panagopoulos:047002} may be of
similar origin to that which we observe.

\section{Experimental procedure}

The polycrystalline samples of La$_2$CuO$_4$ were prepared by a
solid-state reaction using La$_2$O$_3$ and CuO of a minimum purity
of 99.99\%. The respective amounts of the starting reagents were
mixed and then calcinated at 950-1150$^{\circ}$C for 80 hours in
air, with several intermediate grindings. Phase purity of the sample
was checked with X-ray diffractometer (SIEMENS D500).

\begin{figure}
\includegraphics[angle=270,width=8cm]{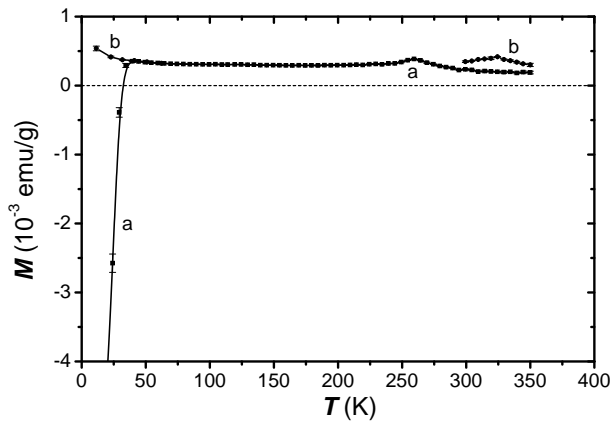}
\caption{\label{Fig1} Temperature dependence of the magnetization of
the as-grown (a) and annealed in argon (b) La$_2$CuO$_4$
policrystalline samples before grinding, $H_{meas} = 1000$ Oe.}
\end{figure}

The as-grown samples were found to be slightly oxygen-enriched
having $T_N \approx 260$ K and containing a very small amount of
superconducting inclusions with $T_c \approx 30$ K and Meissner
phase volume fraction of about $4 \cdot 10^{-4}$. After annealing at
600$^{\circ}$C for 2 hours in a flow of pure Ar ($\sim 99.999\%$)
the sample had $T_N = 325$ K and no detectable diamagnetic
inclusions. In Fig.\ref{Fig1} the temperature dependence of the
magnetization of the sample before and after the annealing in the
measuring field of 1000 Oe is shown.

\begin{figure}
\includegraphics[height=8cm,angle=270]{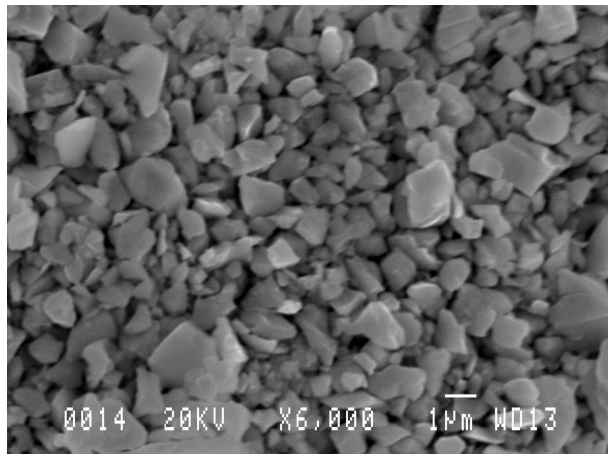}
\caption{\label{Fig2} SEM image of the La$_2$CuO$_4$ sample with the
average grain size of 0.68 $\mu$m.}
\end{figure}

In order to obtain a series of samples with different average grain size,
the annealed compound was ground for 8 hours in an agate
mortar in dry high-purity isopropanol. The ground samples were
dispersed by ultrasound and the grains of different sizes were
extracted by means of successive sedimentation and further
characterized with scanning electron (SEM) and atomic force (AFM)
microscopes. Obtained grain size histograms were well described with
the log-normal distribution. The samples had the grain size
mean values of 0.22, 0.68, 1.53 and 4.1 $\mu$m and the masses were
22, 36, 64 and 35 mg, respectively. The SEM image of the 0.68 $\mu$m
sample is shown in Fig.\ref{Fig2}. X-ray diffraction patterns have
shown that all the samples in the series are crystalline and
single-phase. It is worth noting here that La$_2$CuO$_4$ compound is
stable in air and in water (unlike other cuprates such as
YBa$_2$Cu$_3$O$_{6+y}$ or Sr$_2$CuO$_3$ \cite{Hill_PRB_02}.)

The magnetic measurements were performed with a commercial Quantum
Design MPMS-5 superconducting quantum interference device (SQUID)
magnetometer. The samples were put into the gelatin capsules and
mounted in a polypropylene drinking straw. In order to avoid the
distortion of the $M(H)$ curves due to the observed memory effects,
the "no overshoot" mode for target field approach was used. Scan
length was 4 cm with 24 points per scan. At each point the average
over 5 measurements was taken as a result. The characteristic
measurement time is $\sim 10^2$ s.

\section{Experimental results}

\begin{figure}
\includegraphics[width=8cm]{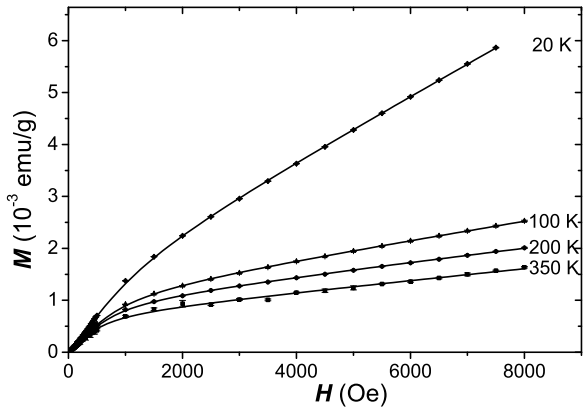}
\caption{\label{Fig3} Temperature variation of the magnetization
curve of the 1.53 $\mu$m La$_2$CuO$_4$ sample. The lines are the
fits with Eq.\ref{Eq0}.}
\end{figure}

\begin{figure}
\includegraphics[width=8cm]{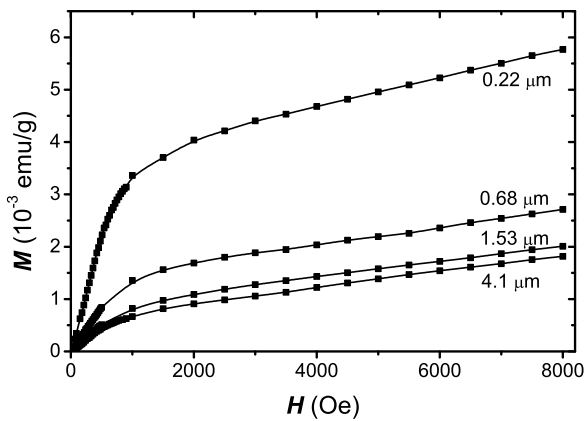}
\caption{\label{Fig4} Magnetization curves of La$_2$CuO$_4$ fine
particle samples for different average grain sizes, $T = 200$ K. The
lines are the guides for an eye.}
\end{figure}

The magnetization curves of the sample with the average grain size
of $\langle d \rangle \approx 1.53$ $\mu$m for different
temperatures are shown in Fig.\ref{Fig3}. In Fig.\ref{Fig4} the
observed at $T = 200$ K curves for different samples are shown. It
is clearly seen that these dependencies consist of two main
contributions: a nonlinear ferromagnetic one, saturating at $H \sim
4000$ Oe, and a linear one. The magnetization curves can be
described thus as
\begin{equation}
\label{Eq0} M(H) = \chi H + M_{nl}(H) .
\end{equation}
The nonlinear component $M_{nl}(H)$ at all the temperatures can be
reasonably well described by the Brillouin (Langevin) function used
for the paramagnetic (superparamagnetic) objects. The fit with the
Brillouin function describing $M_{nl}(H)$ (Fig.~\ref{Fig3}) gives
the temperature-dependent cluster spin values: $11000 \pm 500$,
$5000 \pm 300$, $2000 \pm 200$ and $300 \pm 20$ at 350 K, 200 K, 100
K and 20 K, respectively. The $M(H)$ dependencies for all the
samples were found to be the same, differing only in magnitude
of the linear and the nonlinear terms. The total spin values
characterizing the nonlinear component within the fit error did not
reveal any grain size dependence.

\begin{figure}
\includegraphics[width=7.5cm]{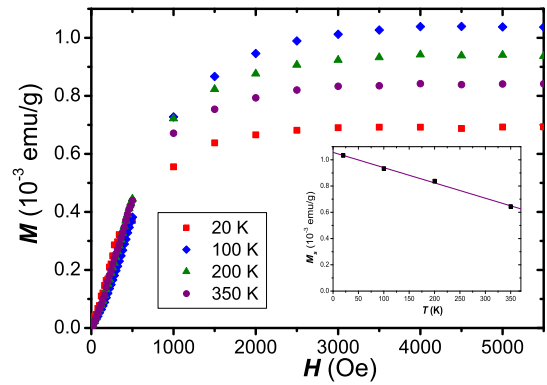}
\caption{\label{Fig5} Temperature variation of the nonlinear
magnetization component of 1.53 $\mu$m La$_2$CuO$_4$ sample. In the
inset the temperature dependence of the nonlinear saturated magnetic
moment is shown.}
\end{figure}

In Fig.~\ref{Fig5} the nonlinear magnetization component $M_{nl}(H)$
obtained by the subtraction of the linear term is shown for a set of
temperature values. This plot indicates that we are not dealing with
the usual superparamagnetic behavior, otherwise the initial slope of
these curves should depend on $T$.

In the inset of Fig.\ref{Fig5} the temperature dependence of the
saturated magnetic moment is shown. In the temperature range $20 -
350$ K it has a nearly linear character.

\begin{figure}
\includegraphics[width=8cm]{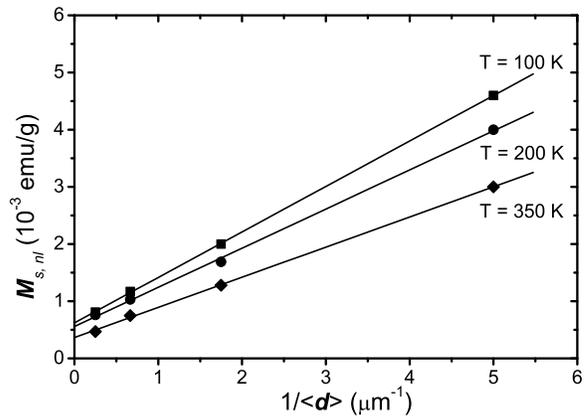}
\caption{\label{Fig6} Grain size dependence of the saturated
nonlinear magnetization component moment.}
\end{figure}

In order to determine if the unusual magnetism of our samples comes
from the bulk or the surface, a grain size dependence of the
magnetization have been studied. In Fig.\ref{Fig6} these data for
the nonlinear magnetization component at several temperatures are
presented. It is characterized by the saturated magnetic moment
$M_s$.

\begin{figure}
\includegraphics[width=8cm]{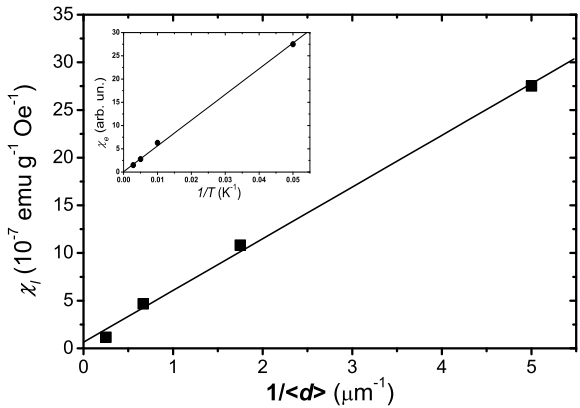}
\caption{\label{Fig7} Grain size dependence of the excess in respect
to the bulk value susceptibility $\chi_e$ describing the linear
magnetization component. In the inset the temperature dependence of
$\chi_e$ is shown for 0.22 $\mu$m La$_2$CuO$_4$ sample.}
\end{figure}

The magnetic susceptibility describing the linear magnetization
component is found to be significantly greater than the
susceptibility of bulk stoichiometric La$_2$CuO$_4$
\cite{Lavrov_PRL03}. Grain size dependence of the {\it excess}
susceptibility with respect to the bulk La$_2$CuO$_4$
$\chi_e=\chi-\chi_B$ at $T = 20$ K is shown in Fig.\ref{Fig7} (this
term is strongly pronounced at low temperatures). The values of
$\chi_B$ for the bulk material were taken from the data of Lavrov et
al. \cite{Lavrov_PRL03} as the orientational averages $\chi_B =
(\chi_a+\chi_b+\chi_c)/3$ (e.g. at $T = 20$ K $\chi_B = 1.81 \cdot
10^{-7}$ emu/(g $\cdot$ Oe)).

The clear linear dependences of both $M_s$ and $\chi_e$ on
$1/\langle d \rangle$, which is the surface to volume ratio, allow
us to assign unambiguously both
 components of $M(H)$ to the grain surface. The
Curie-like $1/T$ dependence of $\chi_e$ shown in the inset of
Fig.\ref{Fig7} reveals a paramagnetic character of this
linear magnetization term.

To investigate the origin of the different components, we annealed
the 0.68 $\mu$m sample after the grinding procedure in the same way
(2 h at 600$^{\circ}$C in the flow of Ar) as it was annealed
initially. It had led to almost total, more than 80\%, removal of
the excess Curie-like component. The nonlinear ferromagnetic
component was reduced much less, only by $\sim 30\%$, indicating
that unlike the linear Curie-like term the latter is rather stable
with respect to annealing.

The explanation of the Curie-like magnetization component now looks
more straightforward. It most probably originates from the surface
Cu$^{2+}$ paramagnetic defects that have been observed in EPR
spectra of La$_2$CuO$_{4+y}$ fine powders and ceramics and described
by W\"{u}bbeler {\it et al} \cite{WubbelerPRB96}. So, in the remainder of the paper
we will be mainly interested in the properties of the nonlinear
magnetization component.

\begin{figure}
\includegraphics[width=8cm]{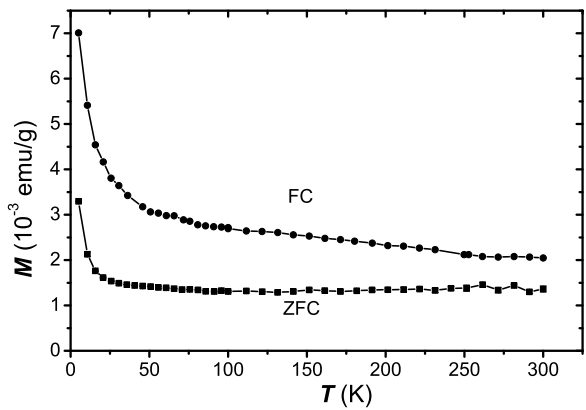}
\caption{\label{Fig8} Temperature dependencies of magnetization of
zero-field cooled (ZFC) and field-cooled (FC) in the field of 10 kOe
for 0.22 $\mu$m La$_2$CuO$_4$ sample.}
\end{figure}

The measured zero-field cooled (ZFC) and field cooled (FC) in the
field of 10 kOe $M(T)$ dependencies are shown in Fig.\ref{Fig8}.
Thermomagnetic irreversibility is found in the whole measured
temperature range $5 - 300$ K and no characteristic spike
corresponding to the blocking temperature is found neither in FC nor
in ZFC curves.

\begin{figure}
\includegraphics[width=8cm]{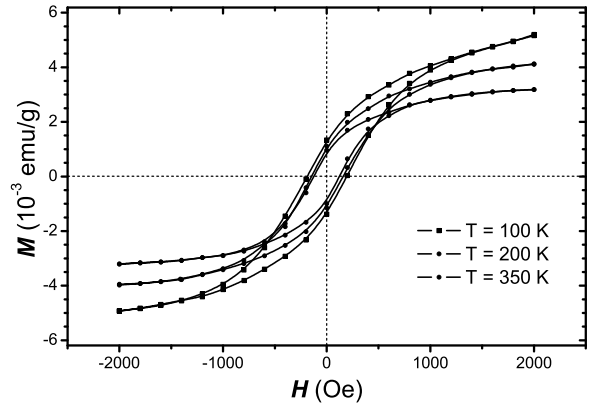}
\caption{\label{Fig9} Magnetization hysteresis loop and its
temperature variation for 0.22 $\mu$m La$_2$CuO$_4$ sample.}
\end{figure}

Hysteresis loops are observed at all temperatures up to 350
K (Fig.\ref{Fig9}). This hysteresis obviously corresponds to the
nonlinear magnetization term. At $T = 100$ K the loop is open up to
~2000 Oe. With increasing temperature the hysteresis loop closes
in gradually decreasing field. This is demonstrated also by the
smoothly decreasing difference between FC and ZFC curves in
Fig.\ref{Fig8}.

The hysteresis loop is symmetric either for FC or for ZFC samples at
any temperature. This means that no detectable exchange bias arising
usually at the boundary of ferromagnetically and
antiferromagnetically ordered phases takes place.

\begin{figure}
\includegraphics[width=8cm]{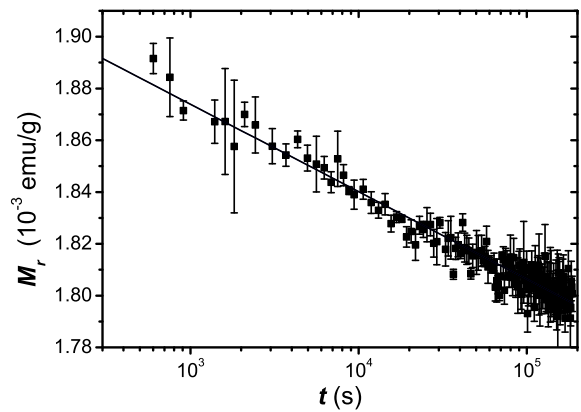}
\caption{\label{Fig10} Decay of the remnant magnetization moment of
the 0.22 $\mu$m sample at $T = 30$ K.}
\end{figure}

The remnant magnetization relaxes rather slowly. The magnetization
decay is shown in Fig.\ref{Fig10}. The decay is described well by
the equation $M(t) = M_0(1 - S \ln t)$, where $M_0$ is the initial
magnetization after the removal of the applied magnetic field and
$S$ is the magnetic viscosity.

A puzzling fact at the first sight is that no characteristic spike
in the $M(T)$ curve corresponding to 3D AF ordering as in
Fig.~\ref{Fig1} was observed for the 0.22 $\mu$m grain size sample
(the data is not shown). In our opinion, this is simply because the
surface magnetization dominated over the bulk one and the dynamic
range of the magnetometer was insufficient to reveal this feature.
Indeed, the spike amplitude for the initial sample in the field of
1000 Oe is $\sim 10^{-4}$ emu/g while the magnetization of the
sample with the grain size 0.22 $\mu$m in this field is $\sim 2.4
\cdot 10^{-3}$ emu/g. The spike thus should be $\sim 4\%$ of the
signal, and taking into account that the sample mass was only 22 mg,
this value could be difficult to detect.

\section{Discussion}

The first question we are going to discuss is whether the observed
magnetization properties are intrinsic for La$_2$CuO$_4$. In our
opinion, usage of high-purity initial components and solvents during
the sample processing allows us to eliminate the possibility of
chemical contamination of the sample. Another possible reason would
be a formation of Cu(OH)$_2$ compound at the surface of the grains
exposed to the air, where the highly reactive broken bonds occur (as
it was mentioned already, La$_2$CuO$_4$ itself is stable in air and
water). Nevertheless, this compound isn't ferromagnetic, so its
formation definitely cannot explain our observations. Also, very
similar magnetic properties were found for CuO
\cite{Punnoose_PRB_CuO} and MnO \cite{Makhlouf_NiO_JAP}
nanoparticles that were synthesized in a totally different way than
in our case. Another argument for an intrinsic origin of our
magnetization is that qualitatively similar rather weak
thermomagnetic irreversibility and hysteresis loops were found even
for high-quality single crystals of La$_{2-x}$Sr$_x$CuO$_4$
\cite{majoros:024528, panagopoulos:047002}, the possible connection
of which to our data will be discussed later.

Our results can be understood well in the following way. The
observation of the hysteresis loops and irreversibility even for the
smallest particles of 0.22 $\mu$m in size in the whole temperature
range 5 - 350 K unambiguously shows that anisotropic ferromagnetic
clusters are formed at the grain's surface. The fit of the nonlinear
magnetization component with the Brillouin function with the total
spin values independent on the grain size within $0.2 - 4$ $\mu$m
range clearly manifests that the average magnetic cluster size is
geometrically significantly smaller than the characteristic size of
the particles. The fit with the Brillouin function should be
nevertheless treated as formal, not revealing any physical
quantities, but demonstrating the universal evolution of the $M(H)$
dependencies with temperature.

\begin{figure}
\includegraphics[width=8cm]{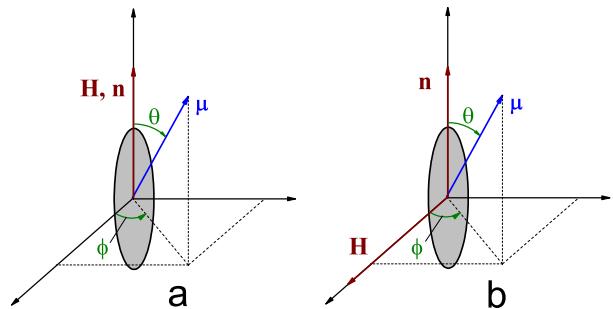}
\caption{\label{Fig11} The sketches used in the calculations of the
ferromagnetic particle energy for the cases of
$\mathbf{H}||\mathbf{n}$ (a) and $\mathbf{H} \bot \mathbf{n}$ (b).}
\end{figure}

In order to explain the details of our observations let us consider
the model describing the magnetization of an ensemble of single-domain
anisotropic ferromagnetic particles (Fig.\ref{Fig11}). We
assume here that local field is  equal to the external field,
implying a low concentration of FM particles. The potential energy of
such a particle in a magnetic field is a sum of the magnetic
anisotropy energy and the energy of interaction of the particle's magnetic
moment with the applied field:
\begin{equation}
\label{Eq1} U = K V \sin^2 \theta - \mathbf{\mu} \cdot \mathbf{H} ,
\end{equation}
where $K$ is the magnetic anisotropy constant, $V$ - the volume of
the particle, $\mu = M_s V$ - the particle magnetic moment, $M_s$ is
the particle saturated magnetization, $\mathbf{H}$ is the applied
magnetic field and $\theta$ - the angle between the particle easy
axis and its magnetic moment. There are two limiting and physically
distinct cases: with the magnetic field applied along and
perpendicular to its easy axis.

In the case of $\mathbf{H}||\mathbf{n}$, where $\mathbf{n}$ is a
unit vector describing the orientation of the particle easy axis
(Fig.\ref{Fig11}, a), the energy of a particle is:
\begin{equation}
\label{Eq2} U = K V \sin^2 \theta - \mu H \cos \theta.
\end{equation}
It can be rewritten introducing the effective anisotropy field $H_a
= 2K/M_s$ and dimensionless field $h = H/H_a = HM_s/2K$ as
\begin{equation}
\label{Eq3} U = \mu H_a \left( \frac{\sin^2 \theta}{2} - h \cos
\theta \right).
\end{equation}

\begin{figure}
\includegraphics[width=7.8cm]{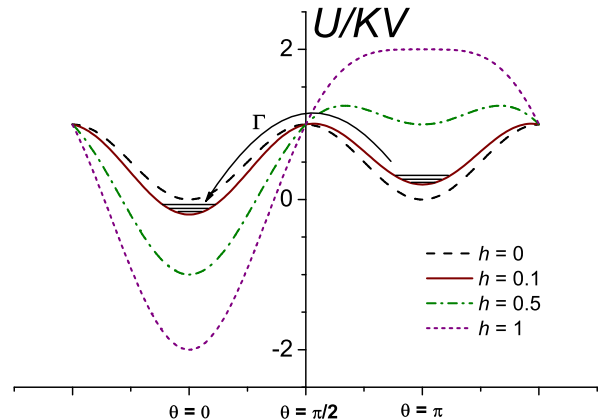}
\caption{\label{Fig11a} Potential energy cross-sections $U(\theta,
\phi)$ for the case of $\mathbf{H}||\mathbf{n}$ with $\phi = 0$ and
different values of applied field $h = H / H_a$.}
\end{figure}

\begin{figure}
\includegraphics[width=7.8cm]{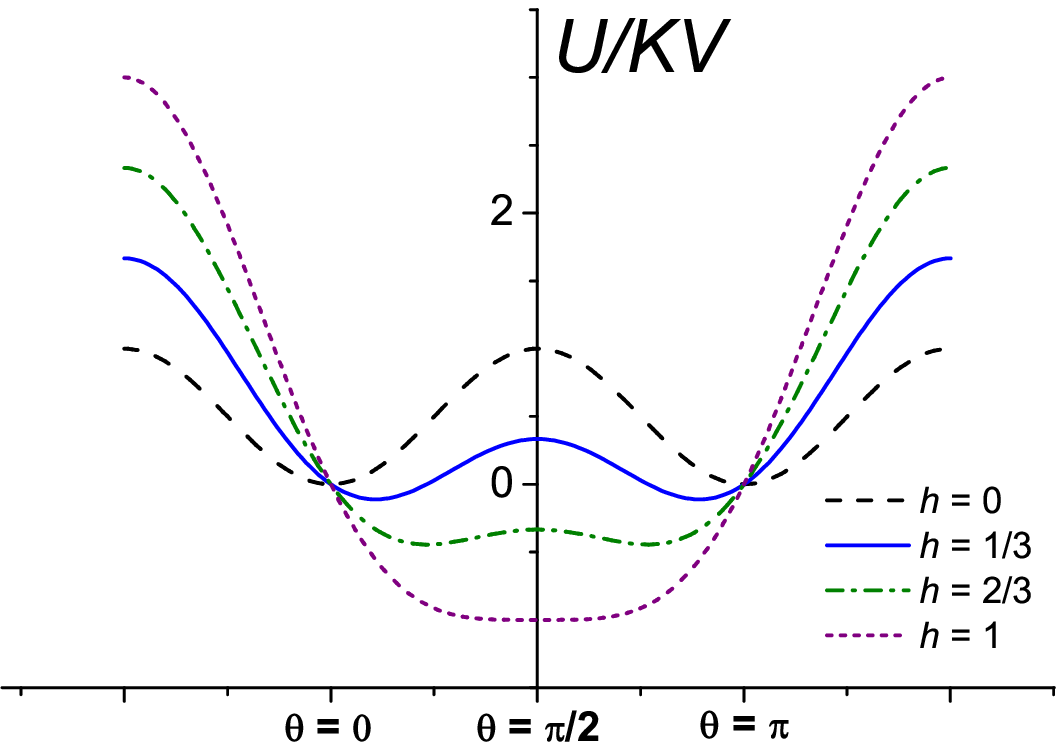}
\caption{\label{Fig12} Potential energy cross-sections $U(\theta,
\phi)$ for the case $\mathbf{H} \bot \mathbf{n}$ with $\phi = 0$ and
different values of applied field $h = H / H_a$.}
\end{figure}

Similarly, for $\mathbf{H} \bot \mathbf{n}$ (Fig.\ref{Fig11}, b),
energy of a particle is described by the expression
\begin{equation}
\label{Eq4} U = \mu H_a \left( \frac{\sin^2 \theta}{2} - h \sin
\theta \cos \phi \right).
\end{equation}
Note that in both cases for the given values of macroscopic
parameters $K$ and $M_s$ the energy landscape amplitude scales with
a particle volume $V$ with the product $\mu H_a = K V$ serving as
scaling factor, while its pattern is defined by the ratio $h =
H/H_a$. The potential energy patterns for these characteristic cases
are shown in Figs.\ref{Fig11a} and \ref{Fig12}.

In both cases two minima separated by the energy barrier are
present, depending on $h$. For $h = 1$ only a single minimum
remains. The principal difference is that with $\mathbf{H} \bot
\mathbf{n}$ energies in the minima are equal, while with $\mathbf{H}
|| \mathbf{n}$ the minima are inequivalent. In the former case the
system is in equilibrium with any applied field. In the latter case
the situation is different, and the relaxation towards equilibrium
is prevented by the small probability to overcome the barrier.

\begin{figure}
\includegraphics[width=7.5cm]{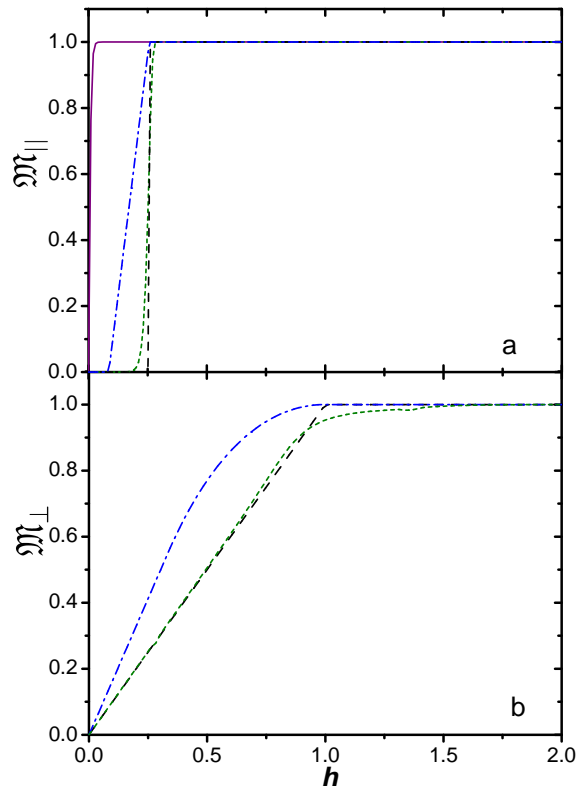}
\caption{\label{Fig14} The simulated magnetization curves for the
particles with $\mathbf{n}||\mathbf{H}$ (a) and $\mathbf{n} \bot
\mathbf{H}$ (b). Short-dash and dashed lines correspond to the
numerical and approximate $M(H)$ curves for fixed $H_a$, while the
dash-dot line is an average for particle collection with the
distribution of $H_a$ as described in the text (field $h$ in this
case is defined as $h = H / H_a^+$); solid line in panel (a) is an
equilibrium magnetization described by Eq.~\ref{Eq6}.}
\end{figure}

Magnetization of the collection of identical particles for these two
cases can be easily treated assuming $k_BT \ll \mu H_a$. So, for
particles with $\mathbf{n} \bot \mathbf{H}$ magnetization takes
place due to the shift of the energy minima towards the direction of
the applied field, and can reasonably well be described as (see also
Fig.~\ref{Fig14}, b)
\begin{equation}
\label{Eq5} \mathfrak{M}_{\bot}(H,T) = \begin{cases}
                                 h&\text{ for $H < H_a$}, \\
                                 1 &\text{ for $H \geqslant H_a$},
                               \end{cases}
\end{equation}
where $\mathfrak{M} = M / M_0$ is a normalized magnetization and
$M_0 = N \mu$ is the total magnetic moment of the system. For
particles with $\mathbf{n} || \mathbf{H}$ the equilibrium
magnetization $\widetilde{\mathfrak{M}}$ within the same assumption
can be approximated by:
\begin{equation}
\label{Eq6} \widetilde{\mathfrak{M}}_{||}(H,T) = \tanh \left( h
\frac{\mu H_a}{k_B T}\right) = \tanh \left( \frac{\mu H}{k_B T}
\right).
\end{equation}
In general for the parallel field the system is only partly in
equilibrium due to the presence of the energy barrier. The measured
magnetization would be a product of equilibrium magnetization to the
fraction of the system that is in equilibrium $\eta(H,T)$:
\begin{equation}
\label{Eq7} \mathfrak{M}(H,T) = \widetilde{\mathfrak{M}}(H,T) \cdot
\eta(H,T).
\end{equation}
This fraction may be easily estimated. The probability $\Gamma$ to
overcome the barrier is defined by N\'{e}el-Brown equation
\begin{equation}
\label{Eq8} \Gamma = \tau^{-1} = \tau_0^{-1} \exp (-U_B/k_B T),
\end{equation}
where $\tau$ is the relaxation time and  $\tau_0 \sim 10^{-10}$ s is
the so-called microscopic attempt time, $E_B$ is the barrier height.
The last is exactly equal to
\begin{equation}
\label{Eq9} U_B(H) = \frac{\mu H_a}{2}\left( 1-h \right)^2,
\end{equation}
and for given characteristic measurement time $\tau_m$ $\eta(H,T)$
is equal to
\begin{equation}
\label{Eq10} \eta(H,T) = \frac{1}{\tau_m}\int_{0}^{\tau_m}
e^{-t/\tau}dt = 1 - e^{-\tau_m/ \tau}.
\end{equation}
If $\tau \ll \tau_m$ at any field, magnetization of such particle
collection will be reversible, and this is the case of
superparamagnetism. In the opposite case, the system is locked in
energy minima until one is destroyed by applied field. Magnetization
of a system then will be irreversible, and this is the case of
ferromagnetism.

The magnetization curves in the conditions of $\mu H_a / k_B T =
0.01$, $\tau_m = 100$ s$^{-1}$ and $\tau_0 = 10^{-10}$ s$^{-1}$ with
the easy axes parallel and perpendicular to the applied field are
shown in Fig.\ref{Fig14}, a and b, respectively.

It is readily seen that $\mathfrak{M}(H)$ curve for the particles
with $\mathbf{n}||\mathbf{H}$ in the limit of $k_B T \ll \mu H_a$
can well be approximated by the theta-function
\begin{equation}
\label{Eq10a} \mathfrak{M}_{||}(H) =
\begin{cases}
  0 &\text{ for $h < h_c$,} \\
  1 &\text{ for $h \geq h_c$,}
\end{cases}
\end{equation}
where $h_c$ is the applied magnetic field, at which $\tau = \tau_m$.
As follows from Eqs.~\ref{Eq8} and \ref{Eq9},
\begin{equation}
\label{Eq10b} h_c = 1 - \sqrt{\dfrac{2 k_B T}{\mu H_a} \ln
\dfrac{\tau_m}{\tau_0}}.
\end{equation}

To make our treatment more realistic we introduce the distribution
of the anisotropy fields $f(H_a)$ in the simplest flat form within
the values $H_a^{-}$ to $H_a^{+}$, leaving other parameters ($K$,
$V$, $\mu$) constant:
\begin{equation}
\label{Eq11} f(H_a) =
\begin{cases}
  \dfrac{1}{H_a^{+}-H_a^{-}} &\text{ for $H_a^{-} \leqslant H_a \leqslant H_a^{+}$,} \\
  0 &\text{ for $H_a < H_a^{-}$ and $ H_a > H_a^{+}$}.
\end{cases}
\end{equation}
From here on there is no sense in using dimensionless field $h$, so
the real one $H$ will be used. For the particles with $\mathbf{n}
\bot \mathbf{H}$ the situation can be treated analytically:
\begin{equation}\label{Eq12}
    \mathfrak{M}_{\bot}(H) = \begin{cases}
             \dfrac{H}{H_a^{+}-H_a^{-}} \ln \dfrac{H_a^{+}}{H_a^{-}} &\text{ for $H \leqslant H_a^{-}$,} \\
             \dfrac{ H - H_a^{-}+H \ln ({H_a^{+}}/{H})}{H_a^{+}-H_a^{-}} &\text{ for $H_a^{-} < H < H_a^{+}$,} \\
             1 &\text{ for $H \geqslant H_a^{+}$.}
           \end{cases}
\end{equation}
At the initial stage the $\mathfrak{M}_{\bot}(H)$ curve is linear
with an effective susceptibility defined by the distribution of
$H_a$. With $H > H_a^-$, the fraction of the system saturates, and
effective slope becomes dependent on $H$, until magnetization is
totally saturated at $H \geqslant H_a^+$. Note, that still for these
particles magnetization process is totally reversible. The
$\mathfrak{M}_{\bot}(H)$ curve for this case with $H_a^+$ and
$H_a^-$ defined by $\mu H_a^+ / k_B T = 100$ and $\mu H_a^- / k_B T
= 300$ is shown in Fig.\ref{Fig14}.

For the particles with $\mathbf{n} || \mathbf{H}$ the situation can
well be approximated if we assume that for all the particles,
similar to the situation described by Eq.~\ref{Eq10a}, the
equilibrium magnetization is saturated at $h > h_c$, which is again
valid for $k_B T \ll \mu H_a$. Thus, we have simply to average
theta-function on the distribution $f(H_a)$. The result is:
\begin{equation}\label{Eq12a}
    \mathfrak{M}_{||}(H) = \begin{cases}
             0 &\text{ for $\widetilde{H} < H_a^{-}$,} \\
             \dfrac{ \widetilde{H} - H_a^{-}}{H_a^{+}-H_a^{-}}
             &\text{ for $H_a^{-} < \widetilde{H} < H_a^{+}$,} \\
             1 &\text{ for $\widetilde{H} \geqslant H_a^{+}$,}
           \end{cases}
\end{equation}
where
\begin{equation}\label{Eq12b}
\widetilde{H} = \dfrac{H}{1 - \sqrt{\frac{k_B T}{K V} \ln
\frac{\tau_m}{\tau_0}}}.
\end{equation}
In this situation $\mathfrak{M}_{||}(H)$ curve reproduces the
dynamics of energy barrier destruction by an applied magnetic field
and thus the derivative $d\mathfrak{M}_{||}/dH$ reflects the
distribution $f(H_a)$ with the renormalized $H_a$ scale. Note that
in this case the magnetization is essentially irreversible. We
should note also that in the real system not only $H_a$ values are
distributed but also particle sizes $V$, magnetic moments $\mu$ and
orientations. Field derivative $dM_{irr}/dH$ would reproduce the
distribution of effective critical fields.

Thus, in the limit of $k_B T \ll \mu H_a$ the magnetization of both
particles with $\mathbf{n} || \mathbf{H}$ and $\mathbf{n} \bot
\mathbf{H}$, only weakly depends on temperature due to small
temperature dependent term in the denominator of Eq.~\ref{Eq12b}.
This explains the minor variation of the field dependence of
magnetization in a wide temperature range $20 - 350$ K
(Fig.~\ref{Fig5}). Field renormalization described by
Eq.~\ref{Eq12b} explains well the gradual decrease of the remnant
moment and coercivity with the temperature increase
(Fig.~\ref{Fig9}).

As far as the magnetization relaxation is concerned, it is clear
that the remnant moment of the monodispersed and aligned particle
collection would be exponential as the energy scale is well defined.
But in a real system presence of the distribution of energy
barriers, as it was shown in Ref.~\onlinecite{Tejada_PRB93}, results
in time-logarithmic decay. It reflects the situation where at any
time of observation $t$, metastable states that are currently
decaying, have the lifetime $\tau \approx t$. This is exactly what
we see in Fig.~\ref{Fig10}.

\begin{figure}
\includegraphics[angle=270,width=8cm]{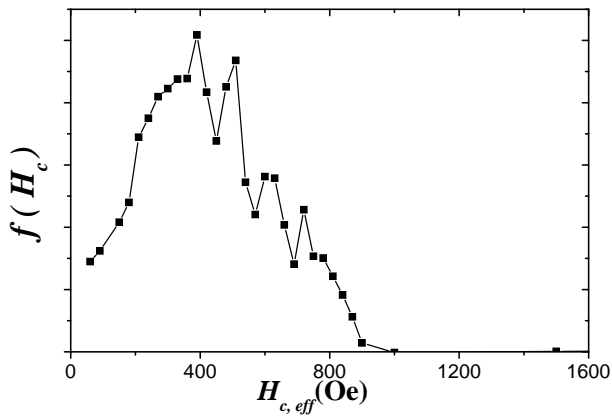}
\caption{\label{Fig15} Distribution of the effective critical fields
obtained from the derivative of the irreversible term of
magnetization curve at $T = 200$ K.}
\end{figure}

Turning to the experimental results, we should note that our object,
which is a collection of ferromagnetic clusters located at the
surface of stoichiometric La$_2$CuO$_4$ grains, is rather poorly
defined both in terms of the cluster size distribution as well as
the values of the constants $M_s$ and $K$. This limits any
quantitative characterization of our sample. Nevertheless some
definite conclusions can be reached. Thus,
\begin{enumerate}
\item[(i)] The fact that surface FM moment scales with the surface to
volume ratio ($\propto 1/ \langle d \rangle$) even for the smallest
$\langle d \rangle = 0.22$ $\mu$m grains (Fig.~\ref{Fig6}) shows
unambiguously that the size of the clusters in a radial direction is
much smaller than $\langle d \rangle / 2 \approx 100$ nm, being of
the order of 10 nm or less. The observation of the universal
magnetization dependencies for all the grain sizes shows that
cluster size distribution is rather universal. Moreover, it probably
indicates that the characteristic size of a cluster is much less
than the grain size for all the samples within the series. Comparing
the observed saturated magnetization of the samples with the
calculated magnetization arising from a surface entirely covered
with a single layer of Cu$^{2+}$ spins, we find the former to be
much smaller, meaning that the clusters are either morphologically
island-like or possibly, cover the entire surface, but have a
weakly-FM canted AF structure. In the latter case, due to the
magnetocrystalline anisotropy we still expect that there would be at
least two domains on each grain, and the model which we have used is
still applicable.
\item[(ii)] The observation of a hysteresis loop at $T = 350$ K, which is well
above $T_N = 325$ K for the bulk compound shows clearly that the
observed surface magnetism is beyond the N\`{e}el hypothesis and is
probably not  due to uncompensated outer planes of the same
sublattice usually observed in binary transition metal oxide
antiferromagnet fine grains. (We consider it unlikely that $T_N$
increases on the surface by 25 K and more in grain samples.)
\item[(iii)] Assuming a homogeneous magnetization of the clusters,
from the data shown in Fig.\ref{Fig5} we can conclude that its
saturated moment $M_s$ decreases almost linearly with temperature in
the range $20 - 350$ K. The observation of a flat ZFC $M(T)$ curve
is the result of peculiar compensation of a drop of $M_s$ with $T$
and the growing fraction of the system in equilibrium.
\item[(iv)] The temperature-independent initial slope of the nonlinear
magnetization component (Fig.~\ref{Fig5}) is determined mainly by
the energy barrier destruction dynamics with the applied magnetic
field. This and the hysteresis loop only slightly depending on $T$
in the range of $5 - 350$ K show that most of the system is in the
limit $k_B T \ll \mu H_a$. Extracting the irreversible part of
magnetization $M_{irr}(H)$ from a difference of a virgin $M(H)$
curve (Fig.\ref{Fig4}) and its reversible part revealed in a
hysteresis loop (Fig.\ref{Fig9}), the distribution of the effective
critical fields $f(H_{c, eff})$ can be estimated as $f(H_{c, eff})
\sim dM_{irr}/dH$, as described above (Fig.\ref{Fig15}). Observation
of the time-logarithmic decay on the time scale of $10^2 - 10^5$ s
at $T = 30$ K (Fig.\ref{Fig10}) together with the hysteresis loops
up to 350 K on the time scale of $10^2$ s (Fig.\ref{Fig9}) allows us
to estimate the energy barrier distribution to be within the range
$0.05 - 1.2$ eV.
\end{enumerate}

Our observation of unusual surface magnetism of fine grains of bulk
antiferromagnet material is not unique. As it has been mentioned
earlier, similar results were found for nanoparticles of most of the
antiferromagnetic transition metal binary oxides. Spontaneous
surface magnetization was observed also in MnF$_2$ single crystal
\cite{Nizhankovskii_EurPhysJB_99, Nizhankovskii_JMMM02}. Close
similarity can be found comparing our data to the results on CuO
\cite{Punnoose_PRB_CuO}, although the authors of
Ref.~\onlinecite{Punnoose_PRB_CuO} hadn't clearly assigned the
ferromagnetic response of their samples to the particles surface. It
looks a bit puzzling that for La$_2$CuO$_4$ fine particles we didn't
observe any exchange bias, like for CuO nanoparticles. This
phenomenon takes place at the interface of FM and AF compounds and
is usually revealed by the shift of the hysteresis loop in field by
some $H_E$ value \cite{MeiklejohnPR57}. One can suggest a number of
possible reasons for this. The magnetic structure of La$_2$CuO$_4$
is more complicated than in binary oxides: it is a 4-sublattice
antiferromagnet with a strong anisotropy. The exchange couplings in
the CuO$_2$ plane are much stronger than the interplane ones. The
in-plane AF correlations appear at much higher temperature than $T_N
= 325$ K, which is a transition to 3D antiferromagnetic state. In
order the exchange bias to be observed one should cool down the
sample in magnetic field through $T_N$, and it is a question which
temperature should be defined as $T_N$ in our case. Another, and in
our opinion the simplest reason for the absence of the exchange bias
can be the fact that our particles are of much greater size than
that for CuO nanoparticles in Ref.~\onlinecite{Punnoose_PRB_CuO}.
Consequently, the volume of the AF core is much larger than the
volume of FM shell, and thus the shell influence is not enough to
stabilize the AF configuration of the core that will define after
the field removal the unidirectional character of the magnetic
anisotropy.

The origin of ferromagnetic response of the antiferromagnetic fine
particles is a matter of current research. N\`{e}el \cite{Neel} was
the first who predicted such a phenomenon, but his hypothesis is not
applicable in our case as it was shown above. Kodama et al.
\cite{Kodama_PRL_NiO} discussed strong coercivity and hysteresis
loop shifts in NiO nanoparticles to arise from the formation of
multisublattice spin configurations due to broken exchange bonds of
the surface sites. Another possible origin of a weak ferromagnetism
at the AF grain surface is the occurrence of Dzyaloshinskii-type
terms \cite{Dzyaloshinskii_JPCS} of the form $\mathbf{D} \cdot [
\mathbf{M_1} \times \mathbf{M_2}]$ in the free energy of the surface
layers; vector $\mathbf{D}$ is normal to the surface, $\mathbf{M_1}$
and $\mathbf{M_2}$ describe the magnetizations of AF sublattices.
This kind of interaction arises due to the loss of inversion
symmetry near the surface. Therefore, finite spin-orbit coupling
would result in spin canting. However, La$_2$CuO$_4$ fine grains are
probably not the best object to study these effects because of its
relative complexity even in the bulk \cite{Kastner_RMP_98,
Lavrov_PRL03}.

Our results seem to be of special importance because La$_2$CuO$_4$
is a parent compound for the high-$T_c$ superconductors. It is now
commonly considered that oxygen and Sr(Ba)-doped La$_2$CuO$_4$ are
strongly inhomogeneous systems. These doped compounds together with
their strong tendency to twin can in principle be treated as the
heterogeneous systems including the AF nearly-stoichiometric grains.
In this sense such a system is similar to our samples and thus the
interface associated magnetism can be found.

Indeed, we mention a number of experimental observations that may be
relevant to our results. Magnetic irreversibilities were observed in
the works by Kremer et al. \cite{Kremer_ZfP_92, Kremer_ZfP_93,
Sigmund_ZfP_94} in phase-separated La$_2$CuO$_{4+y}$ and
La$_{2-x}$Sr$_x$CuO$_4$. Hysteresis loops with the coercivity values
comparable to that in Fig.~\ref{Fig9}, and thermomagnetic
irreversibility were reported by Panagopoulos et al.
\cite{panagopoulos:144508, panagopoulos:047002, majoros:024528} for
La$_{2-x}$Sr$_x$CuO$_4$ in a wide range of Sr concentrations in
polycrystalline and single crystal samples. The presence of local
persistent and superparamagnetically fluctuating magnetic field was
also observed by Chechersky et al. \cite{Chech_PRL93, Chech_PRB96}
by M\"{o}ssbauer spectroscopy in oxygenated and deoxygenated
Nd$_2$CuO$_4$ and superconducting electron-doped
Nd$_{1.85}$Ce$_{0.15}$CuO$_4$ samples. The grain-boundary associated
magnetism has been found by $\mu$SR technique in the
Nd$_{1.85}$Ce$_{0.15}$CuO$_4$ single crystal
\cite{Watanabe_PhysC01}.

According to Ref.~\onlinecite{Nizhankovskii_EurPhysJB_99}, in
MnF$_2$ the surface magnetic moment arises if the dielectric
constant changes significantly at the media boundary, and it even
changes the moment direction for the cases of the $\varepsilon_{in}
> \varepsilon_{out}$ and $\varepsilon_{in} < \varepsilon_{out}$. So,
one of the possible phenomenological reasons of the observed
magnetic clusters formation may be simply a modulation of the
dielectric constant due to inhomogeneities like the mobile hole
segregation that takes place in cuprates.

In conclusion, we have experimentally observed and characterized
surface magnetic phenomena in stoichiometric La$_2$CuO$_4$ fine
grains. The surface gives rise to an excess magnetization with
respect to bulk material, showing both ferromagnetic and
paramagnetic terms in the $M(H)$, both of which scale with the grain
surface area. The paramagnetic term most probably arises from the
surface Cu$^{2+}$ defects reported earlier~\cite{WubbelerPRB96}. For
the ferromagnetic component, hysteresis loops and thermomagnetic
irreversibility are observed in the temperature range $5 - 350$ K,
with remnant moment and coercivity decreasing gradually with $T$.
The observations can be well understood by assuming the formation of
ferromagnetic anisotropic single-domain clusters at the grain
surface. The distribution of effective critical fields is estimated.
The microscopic origin of the ferromagnetic component is tentatively
attributed to symmetry-breaking at the sample surface, but the
detailed origin is unclear at present. The spontaneous appearance of
ferromagnetic islands on the surface of  La$_2$CuO$_4$ might be
useful in nanoscale devices for spin-polarizing the electron current
in multilayer spin valves. The advantage is that antiferromagnetic
oxides are much more common than the ferromagnetic materials
currently in use.
\begin{acknowledgments}
The work was supported within the FP6, project NMP4-CT-2005-517039
(CoMePhS). We thank Z. Jagli\v{c}i\'{c} for his help with the
magnetization measurements.
\end{acknowledgments}

\bibliographystyle{apsrev}
\bibliography{Surf_Magn_LCO_Rev}

\end{document}